\documentstyle[prd,aps]{revtex}
\begin{document}
\tighten
\title{
\begin{flushright}
{\normalsize IIT-HEP-97/3}
\end{flushright}
Remarks on Muon $g-2$ Experiments
and Possible {\em CP} Violation in 
$\pi\to\mu\to e$ Decay}
\bigskip
\author{Daniel M. Kaplan}
\address{Illinois Institute of Technology, Chicago, IL 60616, USA}
\maketitle
\medskip
\centerline{(submitted to Physical Review D)}
\bigskip

While it is generally asumed that $\pi\to\mu\to e$ decay conserves {\em CP}, it
is possible for a small {\em CP}-violating asymmetry to arise from physics
beyond the Standard Model. One possibility to search for such an effect is to
compare the polarization of $\mu^+$ from the decay of $\pi^+$ with that of
$\mu^-$ from the decay of $\pi^-$~\cite{Donoghue}. Such a measurement is
feasible using existing apparatus for the measurement of the anomalous magnetic
moment ($g-2$) of the muon and is sensitive to possible {\em CP} violation both
in pion and in muon decay.

The muon anomalous magnetic moment has been measured to high precision in a
series of experiments at CERN~\cite{Farley-Picasso} and will soon be measured
again at the Brookhaven AGS~\cite{BNL}. In these experiments, which use a 
storage ring to trap muons from $\pi$ decay, electrons from $\mu$ decay are
observed
as the muon spin precesses in a magnetic field. The counting rate of electrons 
oscillates in time, with amplitude $A$ 
proportional to the degree of polarization of the muons,
according to~\cite{Farley-Picasso}
\begin{equation}
N_e(E>E_t,t)=N_0 e^{-t/\tau}[1-A(E_t)\cos{2\pi f_a t+\phi}]\,.
\end{equation}
Here $N_e(E>E_t,t)$ is the number of electrons observed (as a function of the 
time $t$) with energy $E$ 
exceeding a threshold $E_t$, $\tau$ the muon lifetime,
$A(E_t)$ the product of the muon and electron polarizations, 
$f_a$ the precession frequency due to the muon anomalous magnetic moment $g-2$, 
and $\phi$ an arbitrary phase reflecting experimental details.
A precise comparison of the oscillation amplitudes $A_{\pi^+}$ and
$A_{\pi^-}$ for muons from $\pi^+$
and $\pi^-$ decay can thus be extracted from the data.

In Fig.~15 of their paper ``Final Report on the CERN Muon Storage Ring," Bailey
{\it et al.}~\cite{Bailey} give measurements of $A_{\pi^+}$ and $A_{\pi^-}$ for
four threshold energies. While no error estimates are presented, 
they may be derived from the spread in the given values.
The $A(E_t)$ values extracted from that figure are listed in Table 1.
The resulting {\em CP} asymmetry is
\begin{equation}
A_{CP} \equiv \frac{A_{\pi^+}-A_{\pi^-}}{A_{\pi^+}+A_{\pi^-}}\,,
\end{equation}
also shown in Table 1 for each threshold value. Since the four samples are not
independent, the four results should not be averaged. As the threshold is
decreased the polarization decreases, reducing sensitivity to $A_{CP}$. But we
see that for thresholds 1 through 3 the reduction in polarization is
approximately compensated by the increase in statistics. For each
of those 
samples, the limit at 90\% confidence is well approximated by
\begin{equation}
-0.01<A_{CP}<0.02\,,
\end{equation}
which we take as the first available limit on {\em CP} violation in the
$\pi\to\mu\to e$ decay chain.

Since the statistics vary widely for the four samples, with fewer than 10\% of 
events satisfying threshold 1 but 80\% satisfying threshold 4, yet the 
spread of values for each of the four samples is comparable, we may infer that
the measurement is dominated by systematic rather than statistical uncertainty.
The goal of the AGS experiment is to
improve on the CERN results by a factor 20. With attention to the systematics 
of the measurement of $A$, the limit on the {\em CP} asymmetry may improve 
considerably.

{\it Acknowledgements.} The author thanks L. M. Lederman, W.
Morse, and S. Pakvasa for useful discussions.

{\it Note Added.} After submission of this article, the author became aware of
a previous paper by Field, Picasso, and Combley~\cite{Fieldetal} presenting a
similar analysis. While Field, Picasso, and Combley do not proceed as far in
deriving a quantitative result, they present a more detailed discussion of the 
underlying theory.

\begin{table}
\centering
\caption{Oscillation amplitude $A$ (see text) as measured by Bailey {\it et
al.} for four values of the electron energy threshold $E_t$ (extracted from
Fig.~15 of Ref.~\protect\cite{Bailey}) for $\pi^+$ and $\pi^-$ decay, and {\em
CP} asymmetry $A_{CP}$ inferred therefrom.}
\begin{tabular}{l|cccc}
& Th 1 & Th 2 & Th 3 & Th 4 \\
\hline
$A_{\pi^+}$ & $0.385\pm0.003$ & $0.309\pm0.003$ & $0.244\pm0.003$ &
$0.160\pm0.004$\\ 
$A_{\pi^-}$ & $0.383\pm0.005$ & $0.303\pm0.004$ &
$0.238\pm0.001$  & $0.163\pm0.005$\\
\hline
$A_{CP}$ & $0.002\pm0.008$ & $0.009\pm0.009$ & $0.011\pm0.007$ & 
$-0.007\pm0.020$ \\
\end{tabular}
\end{table}

\end{document}